\documentclass[thmsa,a4paper,12pt]{article}
\usepackage{amsfonts}
\usepackage{amssymb}
\usepackage{sw20lart}
\usepackage[doublespacing]{setspace}

\input{tcilatex}
\begin{document}

\title{On Dynamical Cournot Game on a Graph}
\date{}
\author{E.Ahmed\thanks{%
E-mail: magd45@yahoo.com} , M.I.Shehata\footnotemark  \ and H.A.El-Saka%
\thanks{%
E-mail: halaelsaka@yahoo.com} \\
%EndAName
\footnotemark[1] \footnotemark[2] Mathematics Department, Faculty of
Science, Mansoura 35516, EGYPT\\
\footnotemark[3] Damietta Faculty of Science, Mansoura University, 34517,
New Damietta, Egypt\\
}
\maketitle

\begin{abstract}
Cournot dynamical game is studied on a graph. The stability of the system is
studied. Prisoner's dilemma game is used to model natural gas transmission.

\textbf{Key words:}\textsl{\ }\textit{Dynamical Cournot game, graph; gas,
water and electricity transportation.}
\end{abstract}

\section{Introduction}

\bigskip Game theory [1] was first introduced by von Neumann and Morgenstern
in 1944 as a mathematical model. It is the study of ways in which strategic
interactions among rational players produce outcomes with respect to the
preferences of the players. Each player in a game faces a choice among two
or more possible strategies. A strategy is a predetermined program of play
that tells the$i$ player which action to take in response to every possible
strategy other players may use . Transportation of natural gas electricity
and water allocation for different purposes was studied recently by Ilkihc
[2]. He modeled it as static Cournot game on a graph. We study the dynamic
game. The dynamic equations are derived. The stability of the equilibrium
solution is studied. Prisoner's dilemma game is used to model the source
countries-transit countries- markets interaction on a graph.The problem
studied in this letter can be explained by the following example: Assume
Russia wants to export gas to two EU countries say EU1, EU2. Assume
Azerbaijan wants to export gas to EU2 only. Thus Cournot game exists between
Russia and Azerbaijan in EU2. The Russian gas pipelines pass through Belarus
hence prisoner's dilemma game (PD) exists between Russia and Belarus.
Similarly Azeri gas pipelines pass through Turkey hence PD game exists
between Azerbaijan and Turkey.

\section{A Cournot dynamic game on a graph}

Transportation of natural gas electricity and water allocation for different
purposes is an important problem. However most of its studies are static
while the problem is dynamic. The vertices of the graph representing the
problem are pf three types markets m$_{i},i=1,2,...,k1$ firms (producers) $%
f_{j},j=1,2,...,k2$ and transit countries through which the lines pass. The
profit function of firm $f_{j}$ is given by

\begin{eqnarray}
\Pi _{j} &=&\sum_{i}\alpha _{i}q_{ij}-\gamma _{j}s_{j}^{2}/2-\sum_{i}\beta
_{i}q_{ij}c_{i} \\
s_{j} &=&\sum_{l}q_{lj},c_{i}=\sum_{l}q_{il}
\end{eqnarray}%
where $q_{ij}$\ are the production quantity from firm j to market i and $%
\alpha _{i}$, $\beta _{i}$, $\gamma _{j}$ are positive constants. The sum
over i is on all markets connected to firm j. Conversely the sum in $c_{i}$
is over all firms supplying market i.

The dynamic Cournot game with bounded rationality is given by [3,4]

\bigskip 
\begin{equation}
dq_{ij}/dt=b_{j}(\partial \Pi _{j}/\partial q_{ij})
\end{equation}

The parameters $b_{j}$ are proportionality parameters. They may be taken as
functions of the production quantities but here we will take them as
constants.

For general graph Cournot bounded rationality game the dynamic equations are

\bigskip 
\begin{equation}
dq_{ij}/dt=b_{j}[\alpha _{i}-\gamma _{j}\sum_{l}q_{lj}-\beta
_{i}q_{ij}-\beta _{i}\sum_{k}q_{ik}]
\end{equation}

Here we will consider a simple graph consisting of two firms and two
markets. The first firm supplies both markets while second firm supply only
second market. In this case the above system takes the form

\begin{eqnarray}
dq_{11}/dt &=&\alpha _{1}-\gamma _{1}(q_{11}+q_{21})-2\beta _{1}q_{11} \\
dq_{21}/dt &=&\alpha _{2}-\gamma _{1}(q_{11}+q_{21})-\beta
_{2}(2q_{21}+q_{22})  \nonumber \\
dq_{22}/dt &=&\alpha _{2}-\gamma _{2}q_{22}-\beta _{2}(2q_{22}+q_{21}) 
\nonumber
\end{eqnarray}

after rescaling the system becomes:

\begin{eqnarray}
dq_{11}/dt &=&1-r_{1}q_{11}-q_{21} \\
dq_{22}/dt &=&1-r_{2}q_{22}-q_{21}  \nonumber \\
dq_{21}/dt &=&1-r_{3}q_{21}-r_{4}q_{11}-r_{5}q_{22}  \nonumber
\end{eqnarray}

Applying Routh-Hurwitz conditions [5] the unique equilibrium of the above
system is locally asymptotically stable if the following conditions are
satisfied:

\bigskip 
\begin{eqnarray}
a_{1} &>&0,a_{3}>0,a_{1}a_{2}>a_{3} \\
a_{1} &=&r_{1}+r_{2}+r_{3}  \nonumber \\
a_{2} &=&r_{1}r_{2}+r_{1}r_{3}+r_{2}r_{3}-r_{4}-r_{5}  \nonumber \\
a_{3} &=&r_{1}r_{2}r_{3}-r_{1}r_{4}-r_{2}r_{5}  \nonumber
\end{eqnarray}

For the special case $r_{1}=r_{2}$the above results simplifies
significantly. The unique equilibrium solution become

\begin{eqnarray}
q_{11} &=&q_{22}=(1-q_{21})/r_{1} \\
q_{21} &=&(r_{1}-r_{4}-r_{5})/(r_{1}r_{3}-r_{4}-r_{5})  \nonumber
\end{eqnarray}

\bigskip The existence and stability conditions are

\begin{eqnarray}
r_{1} &>&r_{4}+r_{5} \\
r_{3} &>&1  \nonumber
\end{eqnarray}

The approximate solutions for system (6):

If we take $%
\,q_{11}(0)=0.1,q_{22}(0)=0.2,q_{21}(0)=0.3,r_{1}=0.01,r_{2}=0.1,r_{3}=1.1,r_{4}=-0.3, 
$and $r_{5}=0.4.$

We find that the interior equilibrium point is unstable where the conditions
(2.7) is not satisfied.

If we take $%
\,q_{11}(0)=0.1,q_{22}(0)=0.2,q_{21}(0)=0.3,r_{1}=0.2,r_{2}=0.5,r_{3}=1.5,r_{4}=-0.3, 
$and $r_{5}=0.4.$

We find that the interior equilibrium point $(1.13636,0.454545,0.772727)$ is
locally asymptotically stable where the conditions (2.7) is satisfied.

\section{\protect\bigskip Cooperation on graphs}

\bigskip In natural gas, electricity and water transport a new situation
exists namely transit countries. Those where transmission lines pass e.g.
Belarus in the line Russia-Belarus- EU gas line. In this case a Prisoner's
Dilemma (PD) game [6] exists between both producers and end users from one
side and transit countries on the other. In PD game there are two strategies
namely to cooperate or to defect. Typically the payoff matrix for such a
game is given by

\begin{equation}
\left[ 
\begin{array}{cc}
R & S \\ 
T & U%
\end{array}%
\right] ,T>R>U>S
\end{equation}

\ The standard dominant strategy is to defect hence both sides lose a lot. A
way to solve this problems is through side payments paid by both producers
and end users to transit countries.

Games on graphs has been studied by May and Sigmund [7]. They have shown
that in the case of Prisoner's dilemma game cooperation can exist easier
than the standard game.

\section{Conclusions}

Dynamical games representing natural gas, electricity and water are studied
using Cournot game on a graph. Stability of the unique equilibrium solution
is investigated. Prisoner's dilemma game is used to model the producers-
transit countries-end users interactions. The existence of graphs will
improve the possibility of cooperation between these countries.

\end{document}